# GGA-MG: Generative Genetic Algorithm for Music Generation

**Majid Farzaneh**[1] **. Rahil Mahdian Toroghi**[1]
[1]Media Engineering Faculty, Iran Broadcasting University, Tehran, Iran

## Abstract

Music Generation (MG) is an interesting research topic today which connects art and Artificial Intelligence (AI). The purpose is training an artificial composer to generate infinite, fresh, and pleasurable music. Music has different parts such as melody, harmony, and rhythm. In this work, we propose a Generative Genetic Algorithm (GGA) to generate melody automatically. The main GGA uses a Long Short-Term Memory (LSTM) recurrent neural network as its objective function, and the LSTM network has been trained by a bad-to-good spectrum of melodies which has been provided by another GGA with a different objective function. Good melodies have been provided by CAMPIN's collection. We also considered the rhythm in this work and experimental results show that proposed GGA can generate good melodies with natural transition and without any rhythm error.

## 1 Introduction

By growing the size of digital media, video games, TV shows, Internet channels, and home-made video clips, using the music has been grown as well to make them more effective. It is not cost-effective financially and also in time to ask musicians and music composers to produce music for each of the mentioned purposes. So, engineers and researchers are trying to find an efficient way to generate music automatically by machines. Lots of works have proposed good algorithms to generate melody, harmony, and rhythms that most of them are based on Artificial Intelligence (AI) and deep learning.

There are several introduced methods to generate music automatically, such as Hidden Markov Models [21, 5, 19, 7, 16, 20, 2], artificial neural networks [23, 18, 8, 3, 4, 10, 14, 26], the evolutionary and population-based optimization algorithms [13, 25, 11, 22], and local search algorithms [6, 9]. Recently, the sequential deep neural networks especially Long Short-Term Memory (LSTM) neural networks have become prevalently used and achieved successful results generating time series sequences [15, 1, 17].

Music generation can be viewed from different aspects. One can focus on melody generation, while others can work specifically on harmony and rhythm. In terms of data, music generation methods could also be divided into note-based and signal-based methods. In former, the machine should learn music from the music sheets, while in latter the musical audio signals are learned. Music generation can also be discussed in terms of the difficulty of performing [16, 24], and the narrative [12].

In this paper we propose a Generative Genetic Algorithm (GGA) which uses a trained Long Short-Term Memory (LSTM) neural network to generate melodies automatically. This work is an extension of our previous work [32]. In the previous work we used an interactive evolutionary method and only 200 tunes of Campin's collection have been used to train the LSTM network. Also, it was designed to generate melodies in only four common octaves and did not care about the rhythm. But in this work, we have extended the database and octaves and also considered the rhythm. In addition, we have removed the human listeners in this work.

The rest of the paper has organizes as follows:

In section 2, some related works have introduced. In section 3, some key information about the ABC notation language has provided. In section 4, the proposed method has introduced. In section 5, experimental results have reported and finally, in section 6, a conclusion and some suggestions for future researches have provided.

## 2 Related Works

The proposed method uses an evolutionary algorithm to generate melodies. Evolutionary algorithms have extremely used in literature for music generation and other related problems. In this section, some of these works are introduced.

In [27] authors have combined genetic algorithm and genetic programming to create an interactive evolutionary computation based system. This system works for the precise composition of rhythms. By the way, this system requires a human agent to listen to generated rhythms all the time, and it is very time-consuming.

Authors in [28] introduce interactive music composition driven by feature evolution. In this work, Authors use particle swarm optimization and genetic algorithm to examine the convergence behavior of a two-level technique in an objective manner.

In [29] an algorithmic compositional system that uses hill climbing to create short melodies has presented. In this work, the authors use a grammar-based evaluating and similarity with pre-existence melodies to evolve the melodies.

In [30] an evolutionary-based algorithm has proposed which represents musical pieces as a set of constraints changing over time, forming musical contexts allowing to compose, reuse and reshape musical fragments. The system implements a multi-objective optimization aiming for statistical measures and structural features of evolved models.

In [31] a data-based melody generation system has introduced which uses a multi-objective evolutionary computation. In this work, the authors use several evaluation methods simultaneously to calculate the fitness of each generated melody by Non-Dominated Sorting Genetic Algorithm II (NSGA-II).

## 3 ABC Notation

The standard language for this work is the ABC notation language which is very easy to understand and it is extremely useful for us because lots of melodies can be written in a single text *.abc* file and there are lots of functions for working with texts in programming languages.

In this section, we introduce the main commands in ABC notation briefly. For more information, please refer to www.abcnotation.com/learn.

First of all, using some labels, we should determine the general information about our tune. The main labels are X, T, M, L, and K which are tune number, title, meter (which shows the rhythm), default note length, and key signature labels, respectively. For example, the following code in Fig. 1 shows that the piece is number 0, its title is ‘My Tune, the meter is 4/4, default length for each note is ¼ (quarter note), and the key is C. The notes C, D, E, F, G, A, and B has the same characters in ABC notation. For a higher octave, we can use c, d, e, f, g, a, and b characters. For higher octaves, we can use colon (‘) characters, and for lower octaves, we can use comma (,) characters.

Fig. 1 shows a good example of working with notes and octaves.

```
X: 0
T: My Tune
M: 4/4
L: 1/4
K: C
C,D,E,F, | G,A,B,C | DEFG | ABcd | efga |bc’d’e’ | f’g’a’b’ |
```

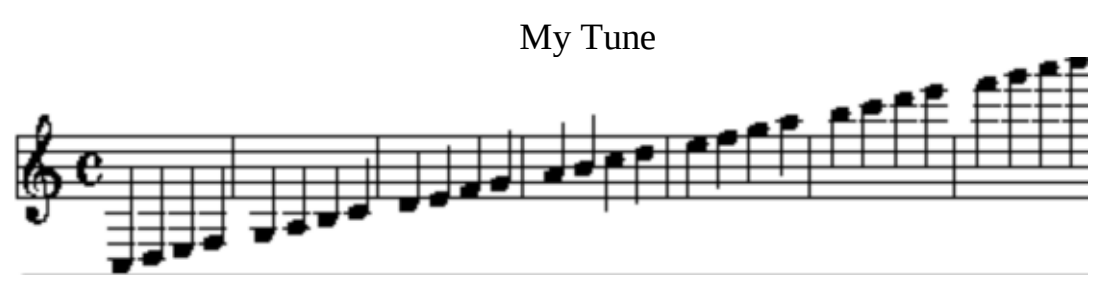

**Fig. 1** An example of ABC notation language

To change note duration, we can simply add a number or a fraction in front of the note character. For example, if note *A* has a default length of ¼ (quarter note), then *A2* has a length of ½ (half note) and *A1/2* has a length of 1/8 (Eighth note), etc.

Accidentals can be added to notes using ‘_’ and ‘^’ characters for flat and sharp, respectively. For natural notes, we can either use ‘=’ character or nothing.

# 4 The Proposed Method

The proposed music generation system has three main phases. At phase I, we use an initial GGA to provide enough sample data for training purposes. At phase II, we use provided training data to train an LSTM neural network, and at phase III, we use the trained LSTM as an objective function for the main GGA. The main GGA is our final music generation system. In this section, we explain the details of the proposed method step by step.

## 4.1 The Main Generative Genetic Algorithm

The Genetic Algorithm in the main GGA tries to generate melodies as its chromosomes and improving them during an optimization process. As we mentioned before, the objective function for this GGA is a previously trained LSTM network. This network receives a melody and provides a score that shows how much the received melody is similar to human melodies. Furthermore, we use a rhythm cost in the objective function that checks how many times the melody breaks the rhythm. So, the main GGA is a two-objective evolutionary optimization algorithm. Fig. 2 shows the main flowchart of the main GGA.

### 4.1.1 Chromosome Structure

Every chromosome in GGAs is a $4 \times N$ matrix that represents a melody, where $N$ is the number of notes in the melody. Each note has four genes including accidental, step, octave, and duration. Table 1 shows the corresponding numbers for ABC notation characters. Fig. 3 shows an example of creating a chromosome from an ABC tune. This structure is used in the LSTM input too. So, the LSTM network has four inputs in each time step.

**Table 1** Chromosome codes for ABC characters

| **Pitch** | | | **Time** |
|---|---|---|---|
| **Accidentals** | **Steps** | **Octaves** | **Durations** |
| ‘_’ → -1 | c C → 1 | A,,, → 1 | Null → 1 |
| ‘^’ → +1 | d D → 2 | A,, → 2 | 2 → 2 |
| ‘=’ or null → 0 | e E → 3 | A, → 3 | 4 → 4 |
| | f F → 4 | A → 4 | 1/2 or / → 0.5 |
| | g G → 5 | a → 5 | 1/4 or // → 0.25 |
| | a A → 6 | a’ → 6 | 1/8 → 0.125 |
| | b B → 7 | a’’ → 7 | 3/2 → 1.5 |
| | z Z → 0 (silent) | | 3/4 → 0.75 |

### 4.1.2 The objective Function

As can be seen in Fig. 2, the objective function consists of two parts, a trained LSTM network, and a rhythm cost calculator. The architecture of the LSTM network is shown in Fig. 4. The Network receives a chromosome as a 4×N matrix which is a sequence and provides a score that shows the fitness of the melody. GGA tries to maximize this score. Also, the rhythm cost should be minimized and in the best situation it should be equal to zero. Rhythm cost is the number of times that melody breaks the rhythm based on the meter. For example, if the meter is 4/4 we need to have a melody with exactly 4/4 time duration in each measure. If by beginning a measure the melody is being in the middle of a note, so the rhythm has been broken. The objective function calculates the fitness using Eq. 1.

$$Fitness = Score_{LSTM} + \frac{\alpha}{\alpha + Cost_{rhythm}} \qquad (1)$$

Where $Score_{LSTM}$ is the output of the LSTM network and $Cost_{rhythm}$ is the number of out of rhythm measures. The constant $\alpha$ is an arbitrary parameter that prevent division by zero, and also can be considered as a regularization parameter between the two objectives.

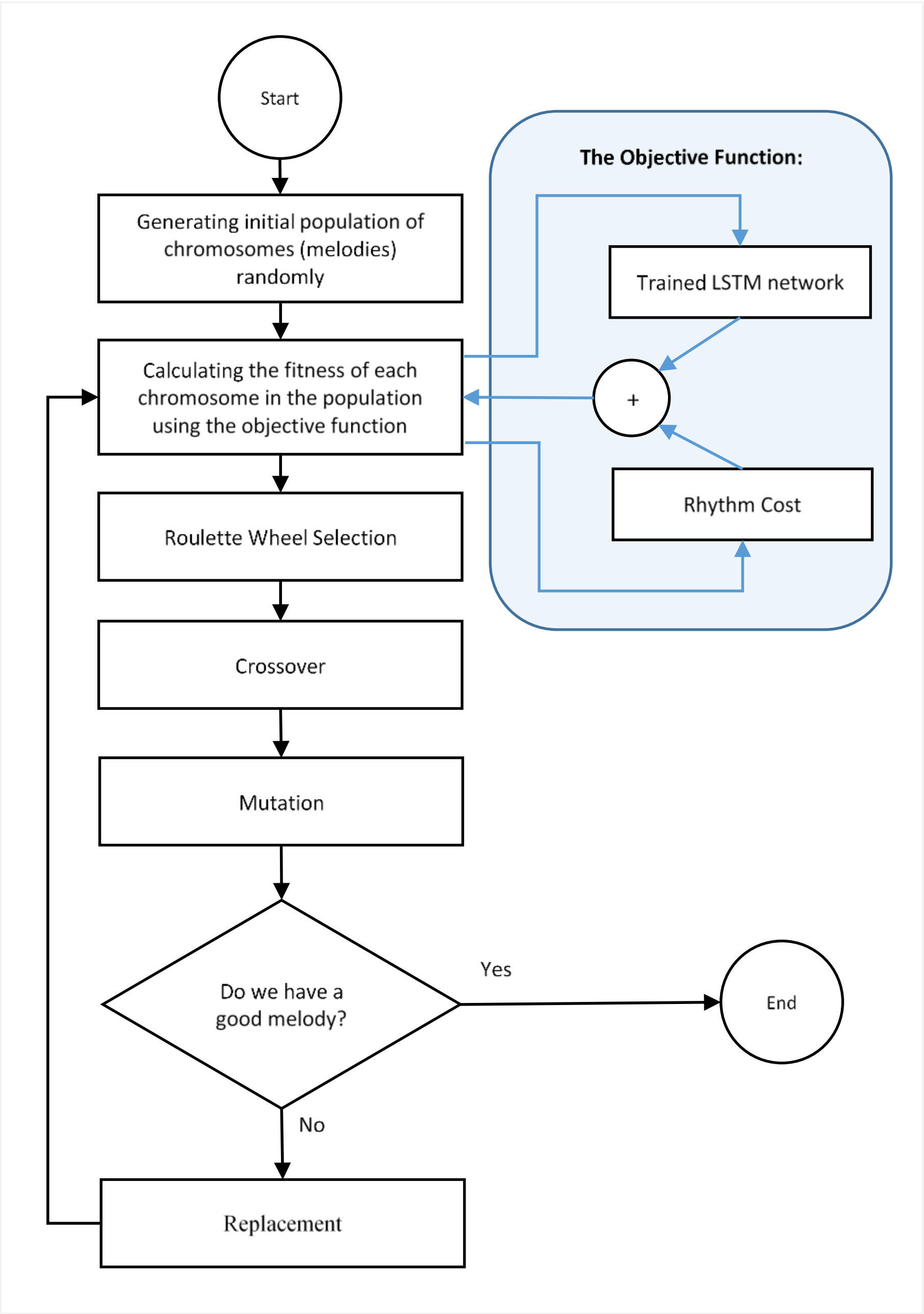

**Fig. 2** The main Generative Genetic Algorithm flowchart

ABC tune: A/ ^B,// _c'3

Corresponding Chromosome:

| | Note 1 | Note 2 | Note 3 |
|---|---|---|---|
| Accidental: | 0 | +1 | -1 |
| Step: | 6 | 7 | 1 |
| Octave: | 0 | -1 | +2 |
| Duration: | 0.5 | 0.25 | 3 |

**Fig. 3** An example of chromosome structure

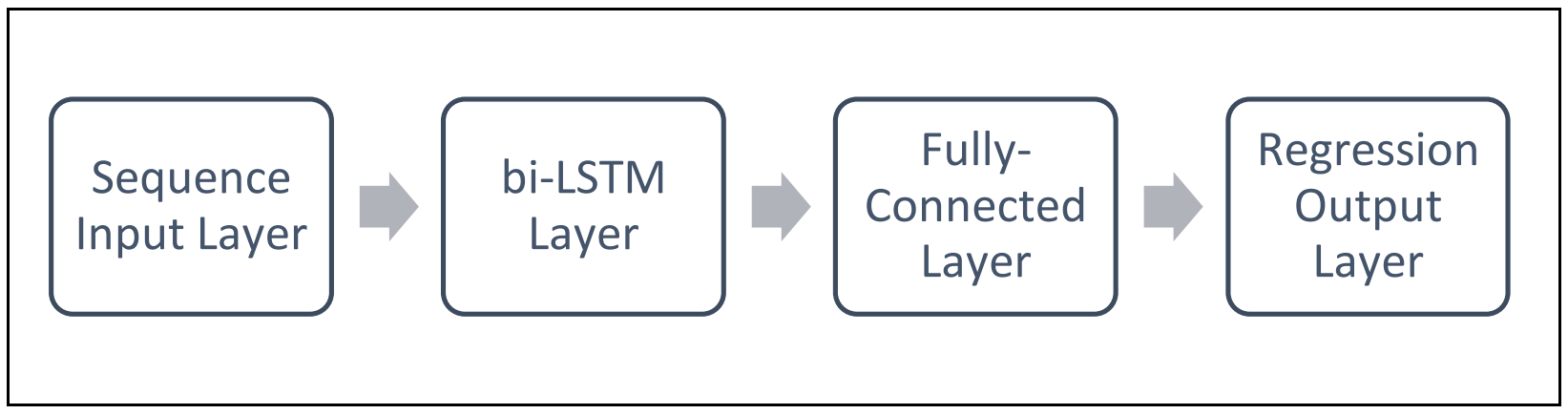

**Fig. 4** The architecture of proposed LSTM network

#### 4.1.3 Initial Population

At the first step of the GGA, an initial population of melodies (chromosomes) should be generated. But in this work we do not generate the melodies completely random. We use the same distribution of the target database in initial population. For this purpose, we calculate the occurrence probability of each character of ABC database, and then we use the probabilities in generation. The character probability can be calculated using Eq. 2:

$$P(character_i) = \frac{N_i}{N_{total}} \qquad (2)$$

Where $N_i$ is the number of occurrence of $character_i$ in the entire database and $N_{total}$ is the total number of characters in the database. Hereafter, whenever GGA generates new melodies, it makes characters occur according to these probabilities. For example, if note G repeated 100 times in database, and there are 5000 characters overall in the database, the probability of character G will be 0.02 and if GGA goes to generate 100 characters for a random melody, 2 characters among those 100 should be G. Therefore, we lead GA makes melodies more like the database.

#### 4.1.4 Selection

In every iteration of the GGA, we need to select some chromosomes as parents for the next generation. For this purpose, we use the Roulette Wheel strategy. In this strategy, in every iteration, first, we calculate a probability for each chromosome based on their fitness values. The probability of selecting chromosome *i* as a parent for next generation, can be calculated using Eq. 3.

$$probability_{selection}(i) = \frac{fitness(i)}{\sum_{j=1}^{Npop} fitness(j)} \qquad (3)$$

By selecting two chromosomes as parents, we can generate two new chromosomes as children for the next generation (iteration). The children can be generated by applying crossover to parents.

To update the population for the next iteration of GGA, we need to repeat selection (based on probabilities), crossover and mutation until we get all required children. Then we replace children with the worth melodies in the population. This strategy is called Roulette Wheel.

#### 4.1.5 Crossover

The crossover is a function that receives two chromosomes, $C_1$ and $C_2$ as parents and generates a new chromosome as a child that contains both $C_1$ and $C_2$ genes. The following pseudo-code shows the proposed crossover function:

```
for i=1 to D do
  if rand>CR then
    child(i)=C1(i);
  else
    child(i)=C2(i);
  end if
end for
```

Where *D* is the number of genes for chromosomes, *rand* is a uniform random number between zero and one, and *CR* is the crossover rate that is a constant number between zero and one and determines how many genes in the child should be from $C_1$ and how many should be from $C_2$. Fig. 5 shows an example of the crossover between two melody matrices.

### 4.1.6 Mutation

To prevent remaining in a locally optimal point, for some generated children we apply a mutation according to the following pseudo-code:

```
for i=1 to D do
    if rand>MR then
        child(i)=child(i);
    else
        child(i)=random_melody(i);
    end if
end for
```

Where *MR* is the mutation rate and determines how many genes in the child chromosome should be replaced by new values. Fig. 5 shows an example of crossover and mutation.

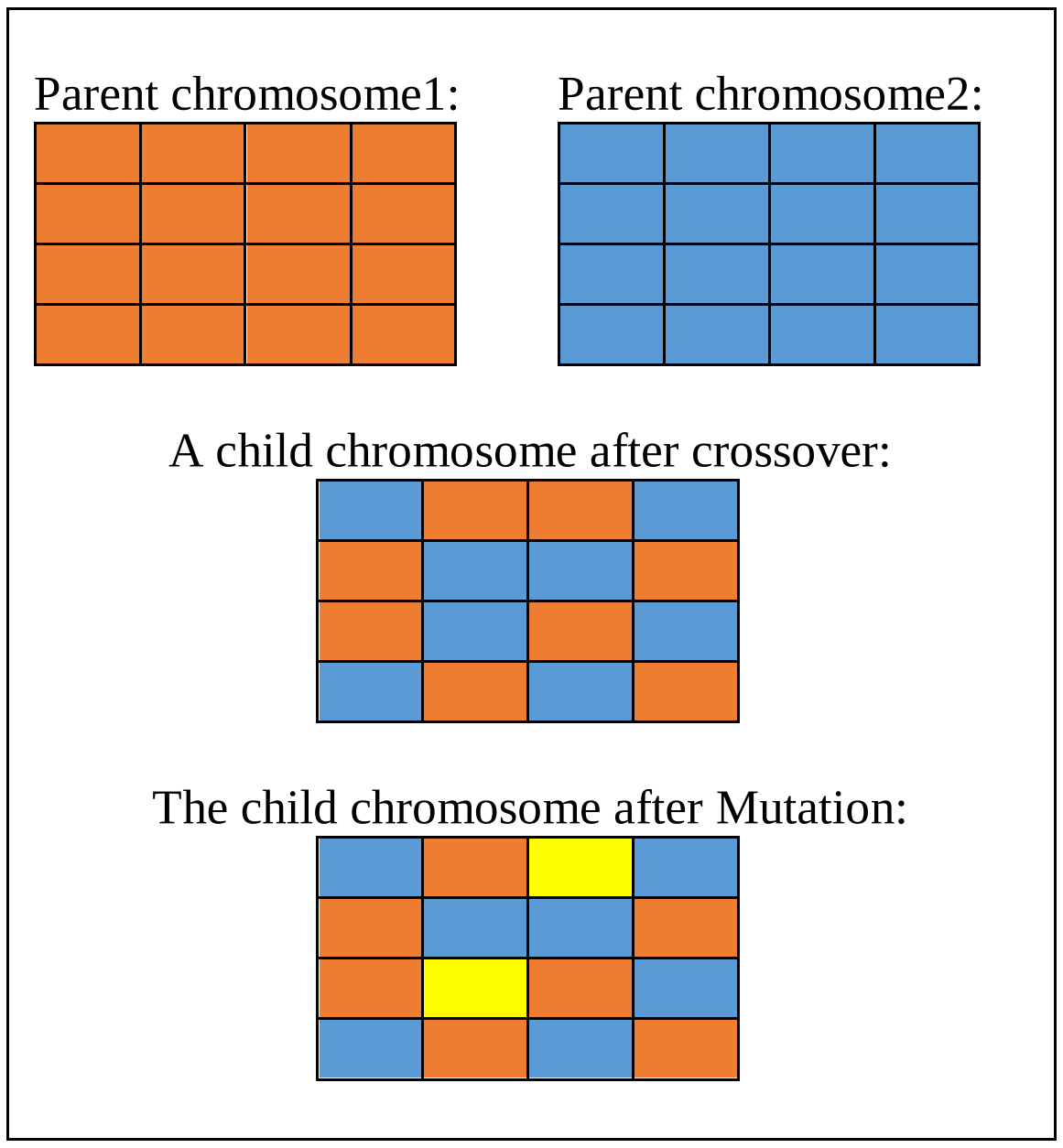

**Fig. 5** An example of crossover and mutation

### 4.1.7 Elitism and Replacement

The population size during the optimization should remain constant unless the execution time will be very costly because of burning new children in every iteration. So, we need to kill some chromosomes in every iteration of the GGAs. But we should transfer only good melodies to the next generation of melodies. This strategy called elitism and it is inspired by natural selection theory by Charles Darwin. So, in every iteration, we pick the best melodies (with the highest amount of fitness values) for the next population and kill others. Instead of poor melodies (melodies with the lowest fitness values), we use the Roulette Wheel strategy and generate some new melodies. According to the above strategy, a population in an iteration contains the best melodies from the previous population and their children.

## 4.2 Long Short-Term Memory

As we mentioned in the sub-section 4.1.2, we use a trained LSTM network as a part of the objective function in the main GGA. The LSTM is a well-known Recurrent Neural Network (RNN) that is very suitable for time-series data. Because it can learn long term dependencies between time samples. Fig. 4 shows a block diagram of the proposed LSTM network. Fig. 6 shows the same architecture in detail. Each LSTM block in Fig. 6 contains several hidden LSTM units. An LSTM unit (also called LSTM cell) usually has a structure similar to Fig. 7 which contains four gates: input, forget, modulation, and output.

We used a bi-directional LSTM (bi-LSTM) layer in our network because in this way it is able to learn the note relations in forward and backward and it is much more effective than a single LSTM. This network should receive a melody and provide a score for it. This score determines how much the input melody is good. The network needs to be trained by several good and bad melodies to learn what a bad melody or a good melody looks like. So, an ideal dataset to train the LSTM network is a Spectrum of bad-to-good melodies. To provide such a spectrum, we use an initial GGN with a different objective function than the main GGN.

**Fig. 6** Detailed Architecture of the proposed LSTM network

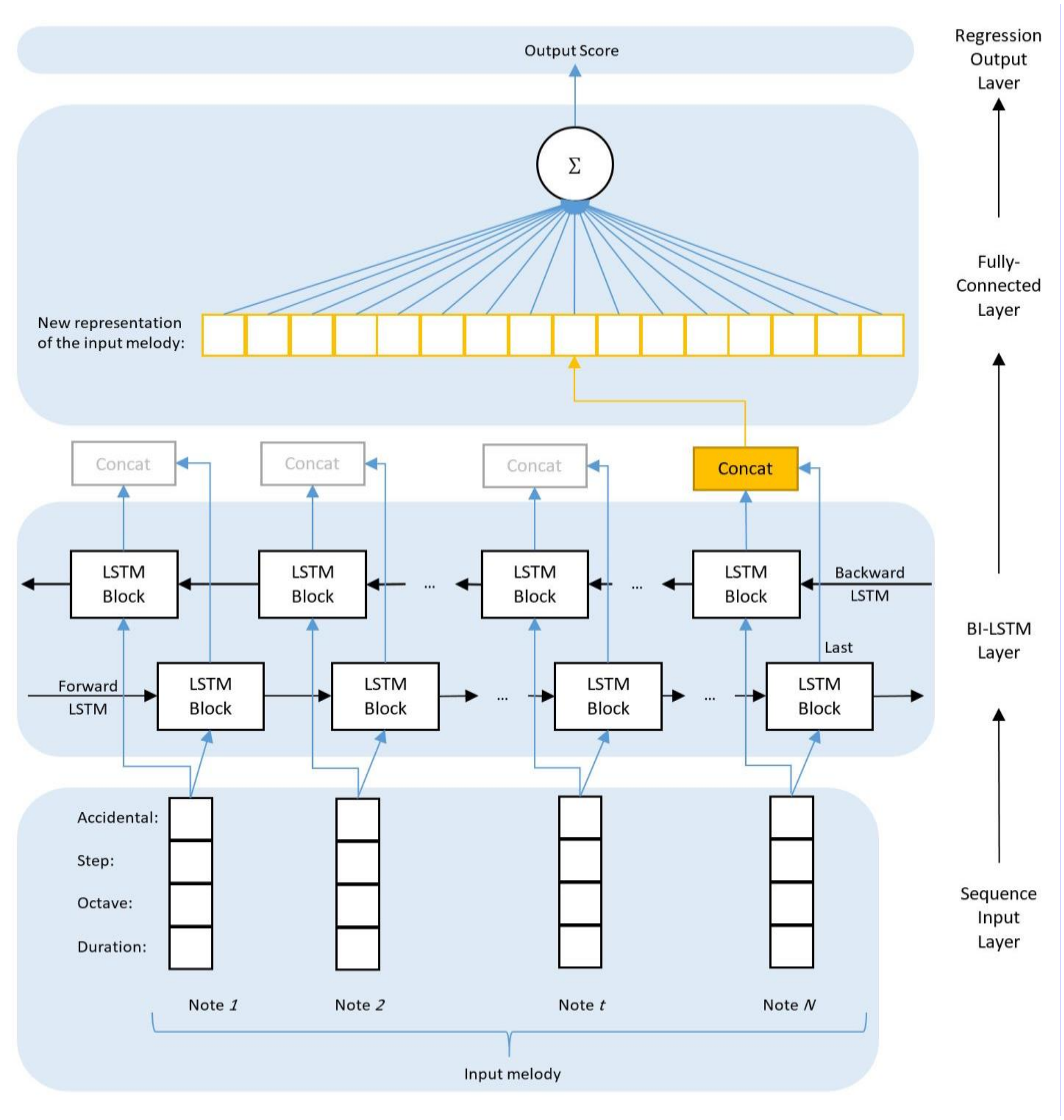

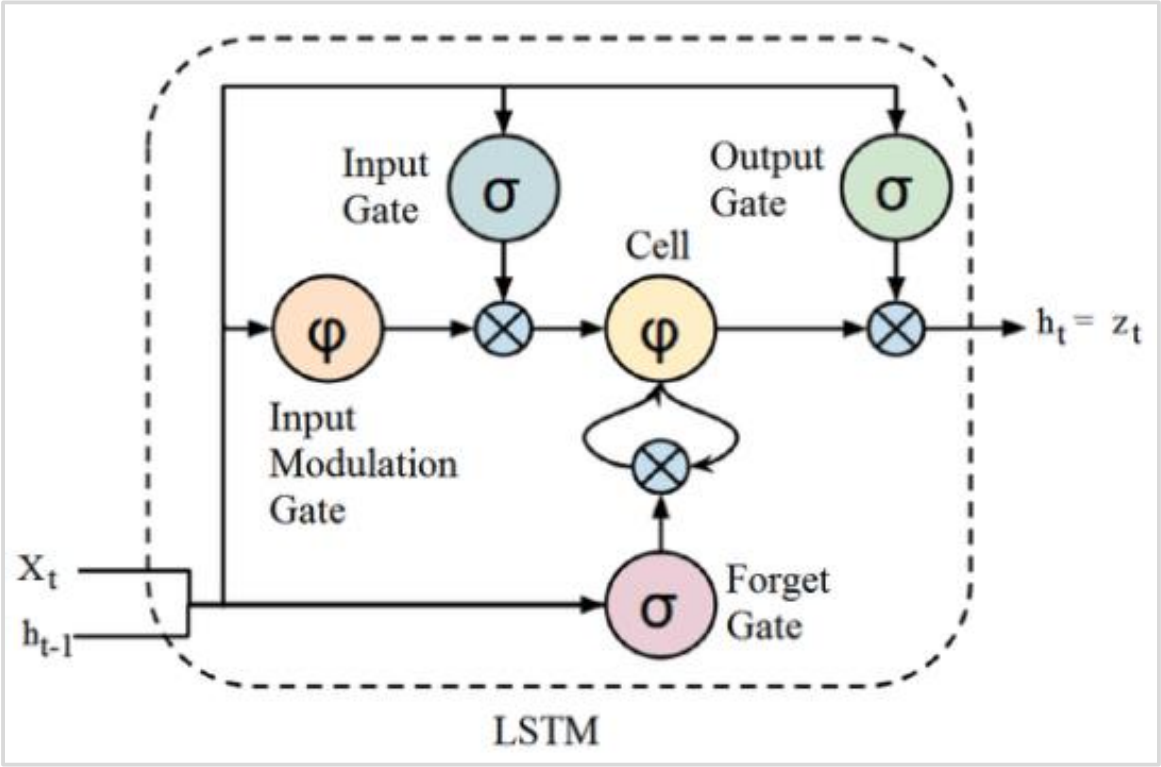

**Fig. 7** LSTM cell structure

## 4.3 Initial Generative Genetic Algorithm

As already mentioned, this phase involves providing the required training data (i.e. melodies) for the LSTM network. Genetic algorithm has been chosen for this task because it can generate a population of melodies (chromosomes) randomly, with a vast spectrum of qualities. These melodies are generated in the same structure of the main GGA.

All parts of the initial GGA are similar to the main GG, but the objective function and mutation are different. Also, unless the main GGA, the initial GGA is not used for music generation. The initial GGA is used to provide *a spectrum* of melodies from a fully-random melody and a good human-made melody. This spectrum might include several melodies from an initial random melody to the optimized one by the initial GGA and this optimized melody is very similar to the target melody. Fig. 8 shows a simple example of a spectrum of random-to-target melodies.

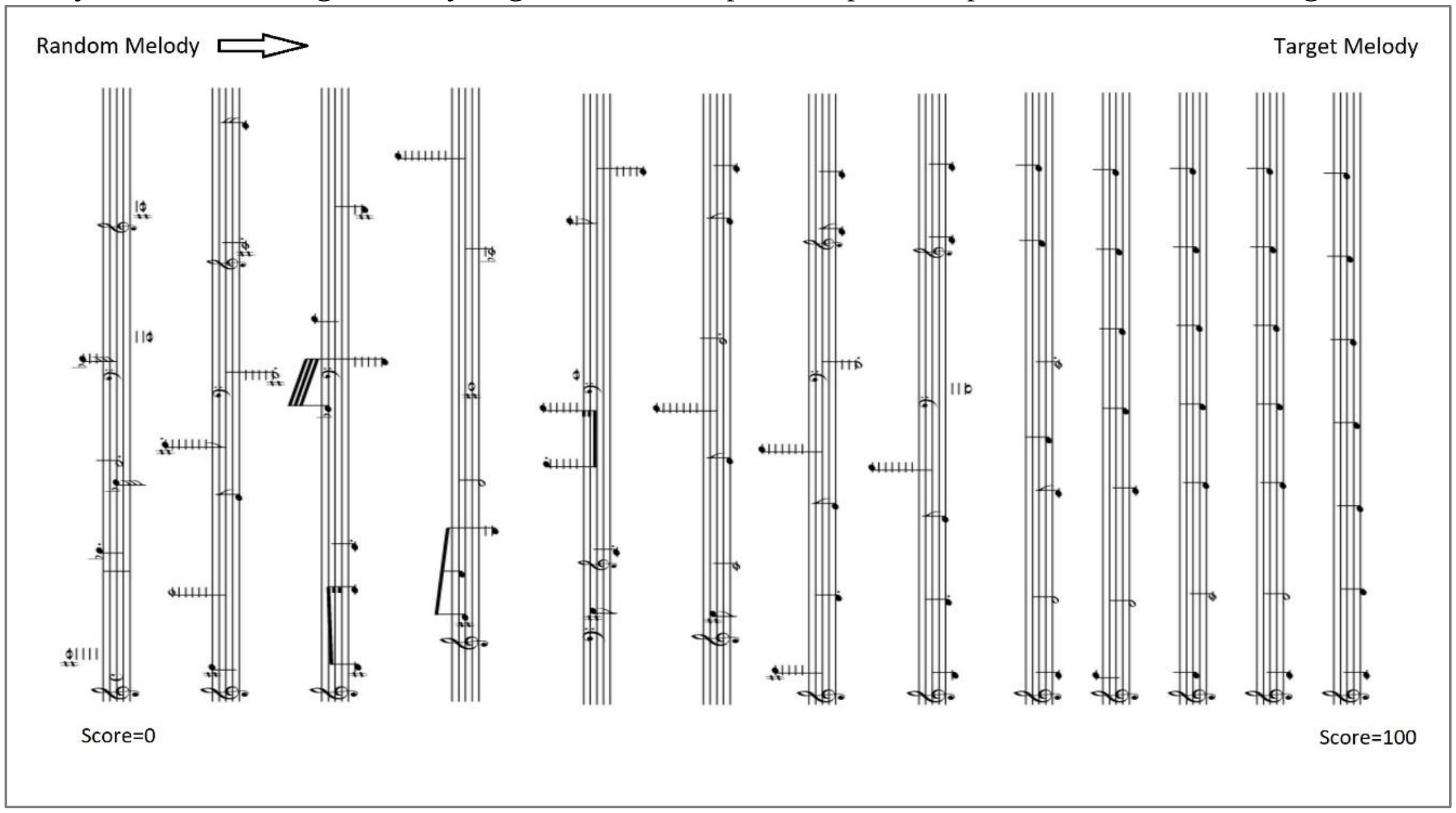

**Fig. 8** *An example of a spectrum of melodies*

To provide required data for the LSTM network, we run the initial GGA several times for different melodies or databases. Every generated melody with its similarity to the human-made melodies is stored as a training sample.

The similarity value is calculated by the objective function of the initial GGA as follows:

$$Fitness(C_i) = \frac{N_2+10N_3+100N_4}{N_i M} \quad (4)$$

Where $C_i$ is i[th] chromosome, $N_i$ is the number of notes of the chromosome $i$, $M$ is the number of notes in human-made melodies, and $N_2$, $N_3$ and $N_4$ are the numbers of 2-grams, 3-grams, and 4-grams which are exist in both generated melody and human-made melodies, respectively.

The initial GGA is an optimization problem that its solution is already known. Because we want to start from a random melody and change it to be an existing melody. So, we use a trick in mutation and instead of changing gens with random values, we change gens with corresponding gens in the target melody. By doing this, the initial GGA will converged much more faster and the required spectrum will be provided. The following pseudo-code shows the mutation in the initial GGA:

```
for i=1 to D do
  if rand>MR then
    child(i)=child(i);
  else
    child(i)=Target_melody(i);
  end if
end for
```

# 5 Experimental Results

## 5.1 Data

The proposed music generation system tries to generate new melodies with high similarity to human-made melodies. So, the proposed algorithm needs to get involved with lots of real melodies. For this purpose we have used the first book of Jack Campin's collection[1] which contains 200 Irish, Scottish and British melodies.

One of the challenges of working with ABC tunes is that one melody can be written in different ways, and the major key and default lengths might be different for different tunes in a specific database. For example, the two tunes in Fig. 9 are equivalent, but melodies have been written differently. To avoid confusing algorithms, we modified the Jack Campin's collection that way the key signature for all tunes have changed to the C key (*K:C*), and the default note has changed to the quarter note (*L:1/4*). Of course, it is not efficient in ABC notation, and it causes longer texts in some tunes, but this is more efficient for the algorithms.

| | |
|---|---|
| X:1<br>T: Bab at the bouster.<br>M:3/4<br>L:1/8<br>K:G<br><br>A/B/c3 B3 \|A BG F2 \| DG AG BA \| BG AG BA \| B2<c2 B3 \|<br>A BA fe \| fg2d ed \| c2<B2 G2 \| A/B/c3 B3 \| | X:2<br>T: Bab at the bouster.<br>M:3/4<br>L:1/4<br>K:E<br><br>A//B//^B3/2 =B3/2 \|A/ B/^^F/ ^F \| =D/^^F/ A/F/ B/A/ \| B/^^F/ A/F/<br>B/A/ \| B<^B =B3/2 \|<br>A/ B/A/ f/e/ \| f/^^f=d/ e/d/ \| ^B<=B ^^F \| A//B//^B3/2 =B3/2 \| |

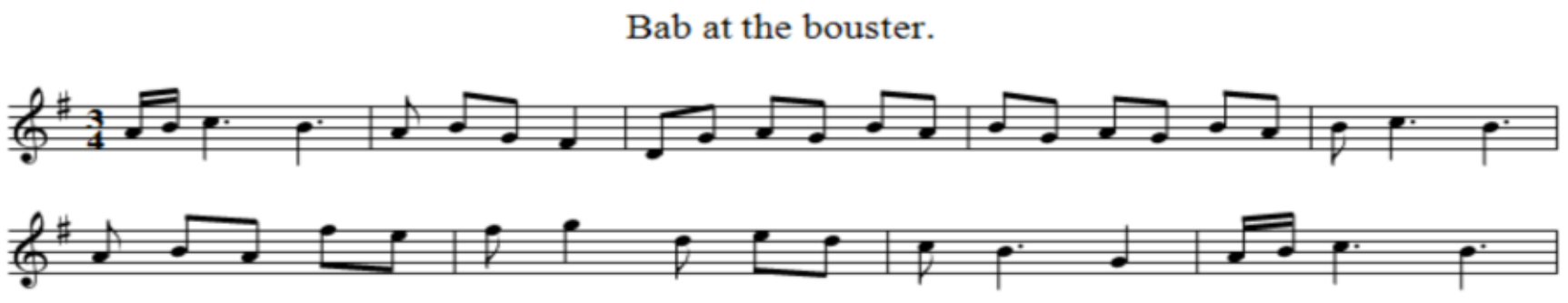

**Fig. 9** An example of two different ways to write a same melody in ABC notation

[1] http://www.campin.me.uk/

Using the initial GGA, we provided 8659 melodies from the Campin's collection. To extend the database we regenerated the 8659 melodies with three different octaves (two octaves lower and one octave higher). Furthermore, we added 4000 fully-random melodies and 4000 random melodies with a similar distribution to the Campin's collection. Totally, our database consist of 42636 different melodies.

All similarity values for the database has been calculated by Eq. 4, but for each target melody, all similarity values have been normalized between 0 and 100. So, the database includes 42636 melodies with different lengths and different scores between 0 and 100, and it used to train the LSTM network.

### 5.2 Settings and Initialization

The proposed method has implemented by MATLAB platform and all runs did by a PC with Intel Core i7 processor, 12 GB of RAM, and Windows 10 OS. Table 2 shows the GGAs and the LSTM parameters for initialization.

**Table 2** GGAs and LSTM parameter values

| Parameter | Value | Explanation |
|---|---|---|
| **POP_Size** | 20 | Population size (number of chromosomes) for both GGAs |
| **MAX_Iter** | 6000 | Maximum iterations for both GGAs |
| **CR** | 0.5 | Crossover rate (half parent1-half parent2) |
| **MP** | 0.5 | Mutation probability (Mutation applies for 50% of children) |
| **MR** | 0.1 | Mutation Rate (10% of genes changes randomly) |
| **Elitism** | 0.5 | 50% of the best chromosomes transfer to the next generation without change |
| **Hidden_Size** | 50 | Number of LSTM units (cells) in the LSTM network |
| **MAX_Epoch** | 200 | Maximum epochs for the LSTM training |
| **Learning rate** | 0.0001 | Learning rate value for the LSTM training algorithm |
| **α** | 10 | Regularization parameter in the objective function of the main GGA |

### 5.3 Results

One of the challenges in the field of music generation is the lack of specific evaluation criterion for the generated music. The LSTM structure, presented in this article can be used as a benchmark if we can trust the network. Fig. 10 shows the proposed LSTM training process for the RMSE and LOSS values of the network.

The data has been divided into 90% for training, and 10% for validation. This figure shows well that the network reached a RMSE of 4.6 after 200 epochs for the validation data. This error may sound large for the regression problems, but in our proposed method it would be acceptable because in any case the main GGA tries to maximize the score from the LSTM structure and the network error in general will have little impact on the music generation process. It even makes the main GGA not generate melodies similar to the dataset. This trivial error, in turn is useful in the sense that it increases the creativity of the system in generating music.

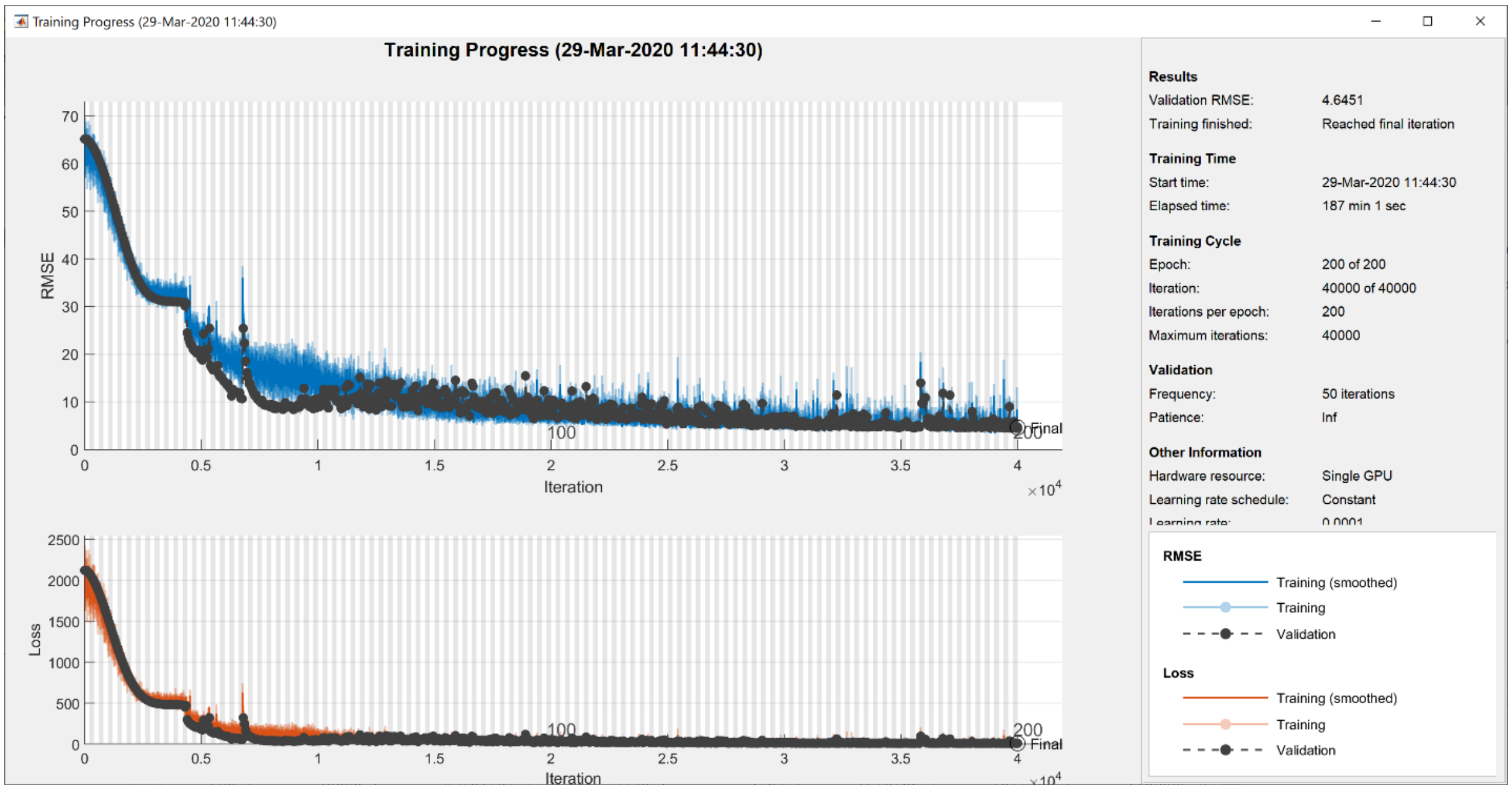

**Fig. 10.** *Training Process of the LSTM network*

Now this question arises whether the proposed LSTM network can effectively help the main GGA generate music? Fig.11 shows four different performances by the main GGA to generate four new melodies of different lengths. As can be seen, the LSTM is remarkably able to assist the main GGA to generate high-rated melodies with scores more than 95. It is also obvious that in only a few primary rounds, the main GGA achieved high scores.

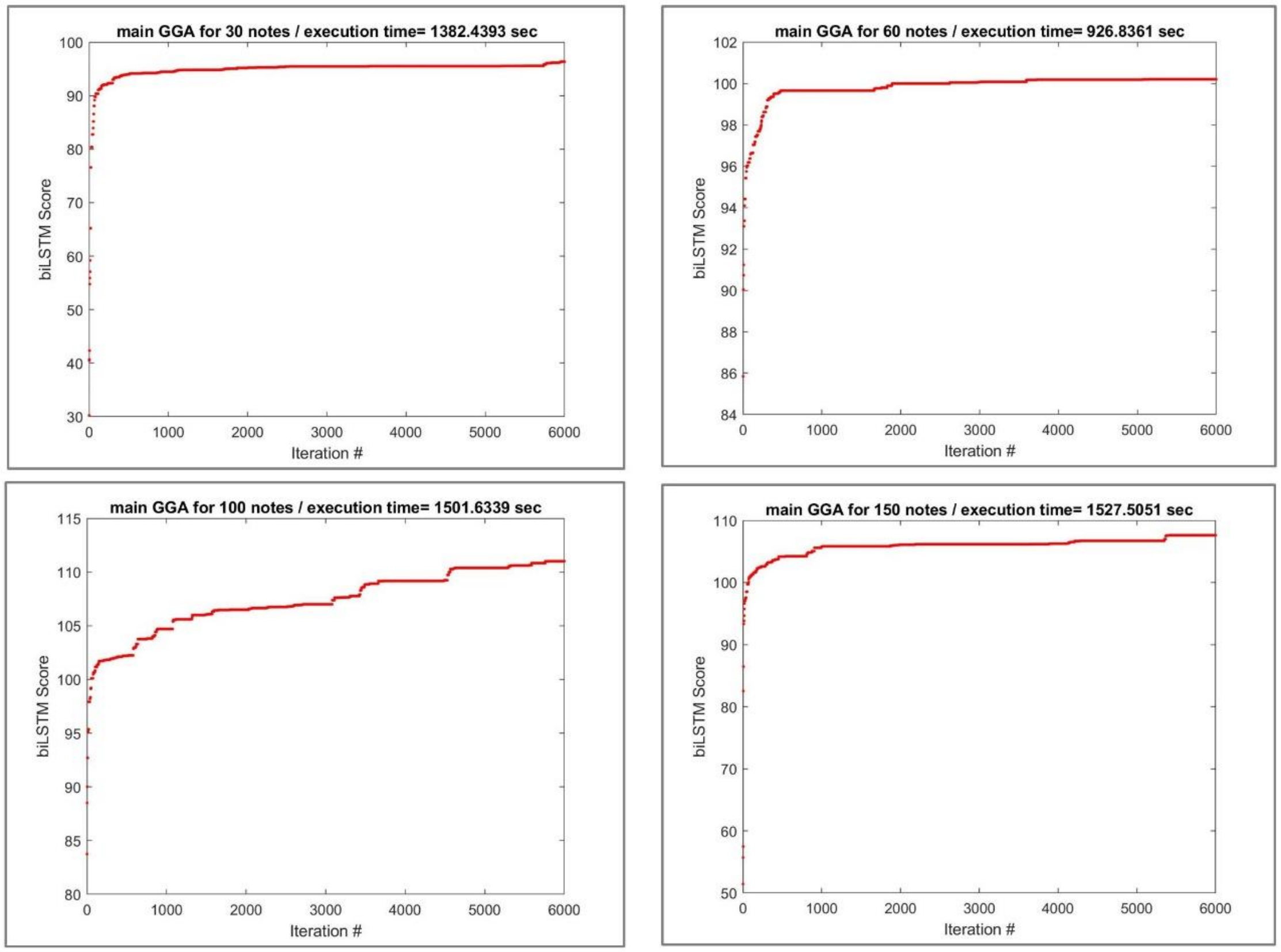

**Fig. 11** *main GGA process for 4 different runs*

We generated four different melodies of the proposed methods using the main GGA and different notes, which could be listened from the link below[2].

These four melodies are evaluated by three different measures and the results are reported in Table3. The evaluations are the LSTM output, the similarity value out of Eq. 4 which has been normalized between zero and 100, and the Mean Opinion Score which involves 10 listeners with scores in the range of 0 to 100, based on how beautiful they feel the melody is.

**Table 3.** *Evaluation*

| **Melody** | **Evaluation** | | |
|---|---|---|---|
| | **LSTM Network** | **Eq. 4** | **Mean Opinion Score** |
| GGA-MG 1 | 107 | 75.5 | 73.2 |
| GGA-MG 2 | 96 | 88.7 | 84.9 |
| GGA-MG 3 | 112 | 84.22 | 61.3 |
| GGA-MG 4 | 108 | 79.53 | 58.0 |

According to the high scores given to the generated melodies, these values indicate that the proposed system is capable of generating acceptable melodies.

While the initial GGA is a time-consuming process, the main GGA system is much faster and produces each note in every 15 seconds using the previously mentioned setup.

Fig. 12 shows the GGA-MG 1 to GGA-MG 4 melodies.

## 6 Conclusion

In this paper, a music generation method based on generative Genetic Algorithm is proposed. A genetic algorithm with the help of an LSTM network generates tunes that are both highly scored and pretty similar to the human tunes. An initial GGA, primarily provides the training data for the LSTM network which has two differences with the standard Genetic algorithms: 1) The optimum solution is already known, 2) The algorithm is not used for optimization, and rather it is used to provide a spectrum of bad-to-good melodies. In fact, in order for the main GGA LSTM network to evolve random melodies into good ones, it needs to be trained for different modes of melodies and this is performed by the initial GGA. The score of the generated melodies is determined by their similarity to a set of human-made melodies. Ultimately, the main GGA optimizes the melodies to get the highest score from the LSTM network, which indicates the highest resemblance to the human-made ones. Creativity is further guaranteed through random generation, LSTM network error and the maximum 4-gram similarity strategy.

Evaluation results indicate that the main GGA is quite eligible to generate melodies more than 70 percent similar to human melodies.

[2] https://soundcloud.com/majid-farzaneh-517982874/sets/generative-genetic-algorithm-music-generator-gga-mg-album

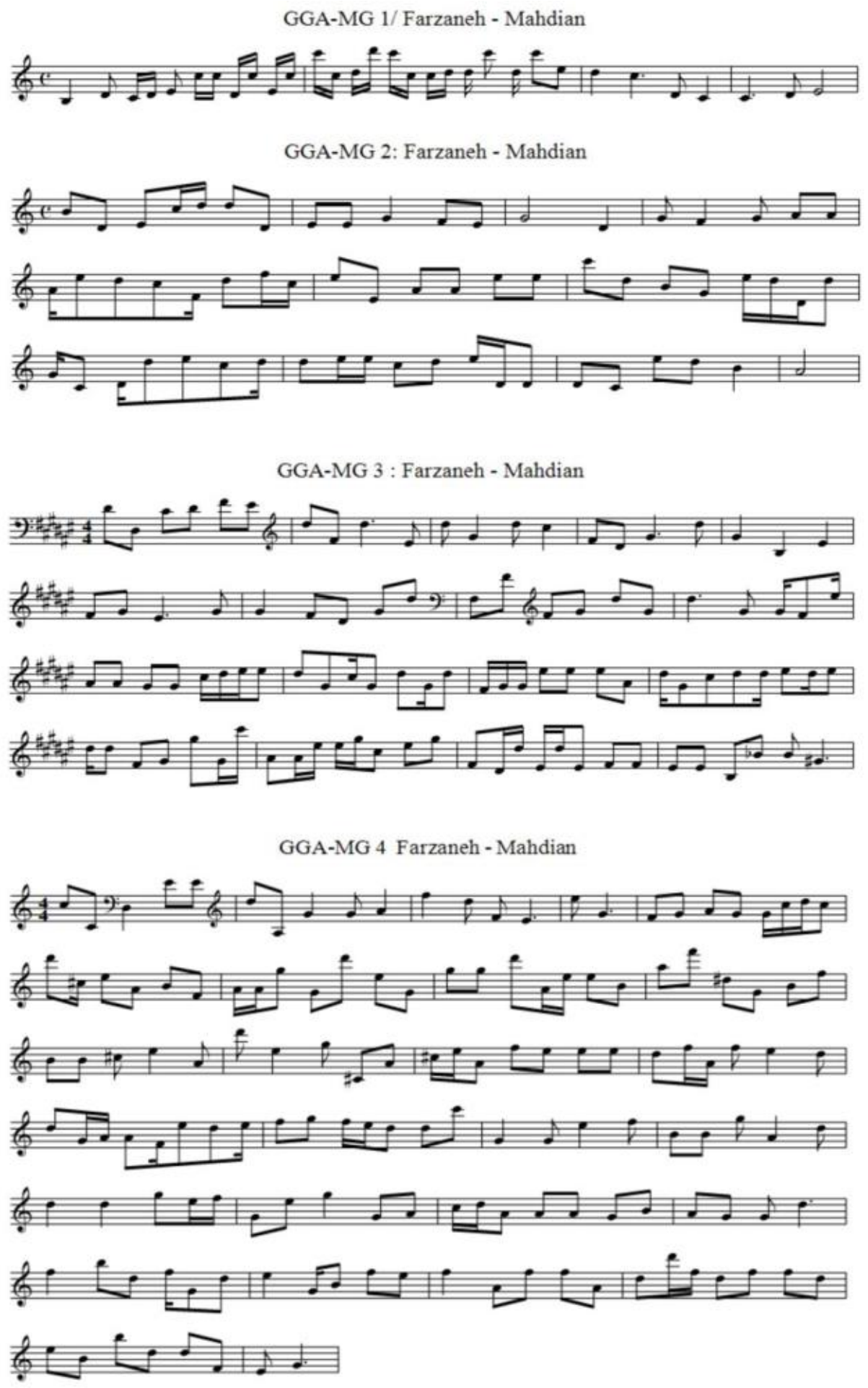

**Fig. 12.** *Sample outputs of the proposed music generation system*

## References

1. Agarwal, S., Saxena, V., Singal, V., Aggarwal, S.: Lstm based music generation with dataset preprocessing and reconstruction techniques. In: 2018 IEEE Symposium Series on Computational Intelligence (SSCI), pp. 455–462. IEEE (2018)
2. Agres, K., Herremans, D., Bigo, L., Conklin, D.: Harmonic structure predicts the enjoyment of uplifting trance music. Frontiers in psychology 7, 1999 (2017)
3. Agres, K.R., DeLong, J.E., Spivey, M.: The sparsity of simple recurrent networks in musical structure learning. In: Proceedings of the Annual Meeting of the Cognitive Science Society, vol. 31 (2009)
4. Boulanger-Lewandowsk, N., YoshuaBengio, P.V., Gray, P., Naguri, C.: Modeling temporal dependencies in high-dimensional sequences
5. Brooks, F.P., Hopkins, A., Neumann, P.G., Wright,W.V.: An experiment in musical composition. IRE Transactions on Electronic Computers (3), 175–182 (1957)
6. Browne, T.M., Fox, C.: Global expectation-violation as fitness function in evolutionary composition. In: Workshops on Applications of Evolutionary Computation, pp. 538–546. Springer (2009)
7. Davismoon, S., Eccles, J.: Combining musical constraints with markov transition probabilities to improve the generation of creative musical structures. In: European Conference on the Applications of Evolutionary Computation, pp. 361–370. Springer (2010)
8. Deb, K., Pratap, A., Agarwal, S., Meyarivan, T.: A fast and elitist multiobjective genetic algorithm: Nsga-ii. IEEE transactions on evolutionary computation 6(2), 182–197 (2002)
9. Eck, D., Schmidhuber, J.: A first look at music composition using lstm recurrent neural networks. Istituto Dalle Molle Di Studi Sull Intelligenza Artificiale 103, 48 (2002)
10. Farzaneh, M., Toroghi, R.M.: Music generation using an interactive evolutionary algorithm. In: Mediterranean Conference on Pattern Recognition and Artificial Intelligence, pp. 207–217. Springer (2019)

11. Herremans, D.: Morpheus: Automatic music generation with recurrent pattern constraints and tension profiles (2016)
12. Herremans, D., Chuan, C.H.: Modeling musical context with word2vec. arXiv preprint arXiv:1706.09088 (2017)
13. Herremans, D., S¨orensen, K.: Composing first species counterpoint with a variable neighborhood search algorithm. Journal of Mathematics and the Arts 6(4), 169–189 (2012)
14. Herremans, D., S¨orensen, K.: Composing fifth species counterpoint music with a variable neighborhood search algorithm. Expert systems with applications 40(16), 6427–6437 (2013)
15. Hofmann, D.M.: A genetic programming approach to generating musical compositions. In: International Conference on Evolutionary and Biologically Inspired Music and Art, pp. 89–100. Springer (2015)
16. Horner, A., Goldberg, D.E.: Genetic algorithms and computer-assisted music composition. In: ICGA, pp. 437–441 (1991)
17. Huang, Z., Xu, W., Yu, K.: Bidirectional lstm-crf models for sequence tagging. arXiv preprint arXiv:1508.01991 (2015)
18. Kaliakatsos-Papakostas, M.A., Floros, A., Vrahatis, M.N.: Interactive music composition driven by feature evolution. SpringerPlus 5(1), 826 (2016)
19. Ponce de Le´on, P.J., I˜nesta, J.M., Calvo-Zaragoza, J., Rizo, D.: Data-based melody generation through multi-objective evolutionary computation. Journal of Mathematics and Music 10(2), 173–192 (2016)
20. Lipowski, A., Lipowska, D.: Roulette-wheel selection via stochastic acceptance. Physica A: Statistical Mechanics and its Applications 391(6), 2193–2196 (2012)
21. Loughran, R., McDermott, J., O'Neill, M.: Grammatical music composition with dissimilarity driven hill climbing. In: International Conference on Computational Intelligence in Music, Sound, Art and Design, pp. 110–125. Springer (2016)
22. Makris, D., Kaliakatsos-Papakostas, M., Karydis, I., Kermanidis, K.L.: Combining lstm and feed forward neural networks for conditional rhythm composition. In: International Conference on Engineering Applications of Neural Networks, pp. 570–582. Springer (2017)
23. Manzelli, R., Thakkar, V., Siahkamari, A., Kulis, B.: An end to end model for automatic music generation: Combining deep raw and symbolic audio networks. In: Proceedings of the Musical Metacreation Workshop at 9th International Conference on Computational Creativity, Salamanca, Spain (2018)
24. McVicar, M., Fukayama, S., Goto, M.: Autoleadguitar: Automatic generation of guitar solo phrases in the tablature space. In: 2014 12th International Conference on Signal Processing (ICSP), pp. 599–604. IEEE (2014)
25. Mishra, A., Tripathi, K., Gupta, L., Singh, K.P.: Long short-term memory recurrent neural network architectures for melody generation. In: Soft Computing for Problem Solving, pp. 41–55. Springer (2019)
26. Music, A.: Creation by refinement and the problem of algorithmic music composition. Music and connectionism p. 212 (1991)
27. Pachet, F., Roy, P., Barbieri, G.: Finite-length markov processes with constraints. In: Twenty-Second International Joint Conference on Artificial Intelligence (2011)
28. Papadopoulos, A., Roy, P., Pachet, F.: Avoiding plagiarism in markov sequence generation. In: Twenty-Eighth AAAI Conference on Artificial Intelligence (2014)
29. Pinkerton, R.C.: Information theory and melody. Scientific American 194(2), 77–87 (1956)
30. Scirea, M., Togelius, J., Eklund, P., Risi, S.: Affective evolutionary music composition with metacompose. Genetic Programming and Evolvable Machines 18(4), 433–465 (2017)
31. Todd, P.M.: A connectionist approach to algorithmic composition. Computer Music Journal 13(4), 27–43 (1989)
32. Tokui, N., Iba, H., et al.: Music composition with interactive evolutionary computation. In: Proceedings of the third international conference on generative art, vol. 17, pp. 215–226 (2000)
33. Tuohy, D.R., Potter, W.D.: A genetic algorithm for the automatic generation of playable guitar tablature. In: ICMC, pp. 499–502 (2005)
34. WASCHKA II, R.: Composing with genetic algorithms: Gendash. In: Evolutionary Computer Music, pp. 117–136. Springer (2007)
35. Wu, J., Hu, C.,Wang, Y., Hu, X., Zhu, J.: A hierarchical recurrent neural network for symbolic melody generation. IEEE Transactions on Cybernetics (2019)